\documentclass[conference]{IEEEtran}
\usepackage[left=0.75in,right=0.75in,bottom=0.75in,top=0.75in]{geometry}  
\IEEEoverridecommandlockouts                    \def\IEEEtitletopspace{0.25in}
\usepackage{subcaption}
\usepackage{graphicx} 
\usepackage{balance}
\usepackage{comment}
\usepackage{cite}
\usepackage{amssymb}
\usepackage[active]{srcltx}
\usepackage{amsmath}
\usepackage{amsfonts}

\newcommand{\norm}[1]{\left\lVert#1\right\rVert}
\graphicspath{{./figs/}}
\usepackage{color,soul}
\usepackage{eurosym}
\usepackage{algorithm}
\usepackage{algorithmic}
\usepackage{multicol}
\usepackage{xcolor}
\usepackage{multirow}
\usepackage{footnote} 
\usepackage[hidelinks]{hyperref}
\floatname{algorithm}{Algorithm}

\begin{document}
\title{Cooperative Search and Track of Rogue Drones using Multiagent Reinforcement Learning}

\author{Panayiota Valianti, Kleanthis Malialis, Panayiotis Kolios, and Georgios Ellinas
\thanks{P. Valianti, K. Malialis, and G. Ellinas are with the Department of Electrical and Computer Engineering and the KIOS Research and Innovation Center of Excellence (KIOS CoE), University of Cyprus, Nicosia, 1678, Cyprus. P. Kolios is with the Department of Computer Science and the KIOS CoE, University of Cyprus, Nicosia, 1678, Cyprus. E-mail: \{pvalia01, kmalia01, pkolios, gellinas\}@ucy.ac.cy. \newline
This work was supported by the European Union's Horizon 2020 research and innovation programme under grant agreement No 739551 (KIOS CoE - TEAMING) and from the Republic of Cyprus through the Deputy Ministry of Research, Innovation and Digital Policy.}}

\maketitle

\begin{abstract}
This work considers the problem of intercepting rogue drones targeting sensitive critical infrastructure facilities. While current interception technologies focus mainly on the jamming/spoofing tasks, the challenges of effectively locating and tracking rogue drones have not received adequate attention. Solving this  problem and integrating with recently proposed interception techniques will enable a holistic system that can reliably detect, track, and neutralize rogue drones. Specifically, this work considers a team of pursuer UAVs that can search, detect, and track multiple rogue drones over a sensitive facility. The joint search and track problem is addressed through a novel multiagent reinforcement learning scheme to optimize the agent mobility control actions that maximize the number of rogue drones detected and tracked. The performance of the proposed system is investigated under realistic settings through extensive simulation experiments with varying number of agents demonstrating both its performance and scalability.
\end{abstract}


\section{Introduction} \label{sec:intro}
The rapid technological progress in unmanned aerial vehicles (UAVs) and the surging global demand have opened up a wide range of applications for the use of UAVs in both the military and more importantly the civilian sectors \cite{9194729}. However, the simplicity of operating UAVs poses significant security, and public safety risks \cite{8678424}. The unauthorized or illegal use of drones over sensitive, restricted, and critical infrastructures necessitates the development of robust and precise counter-drone solutions that can be effective in intercepting rogue drones \cite{hackingdrones}.

Evidently, current interception technologies are still in their infancy, with only a handful of systems capable of effectively neutralizing rogue drones. The most popular countermeasures include the deployment of radio frequency (RF) jammers that can disrupt a drone’s GNSS (localization/navigation) or RF signals (mostly used for telemetry) \cite{9378538}. Nevertheless, the interference levels caused by these jamming systems pose serious side effects to the normal operation of authorized systems operating within sensitive areas \cite{FAA}, \cite{9765451}.

Moreover, the interception task is only part of a multifaceted and complex problem of detecting, tracking, and neutralizing rogue drones, due to the underlying challenges entailed by the dynamic behavior of heterogeneous drones, accuracy and efficiency of detection systems, and flexibility and scalability of neutralization systems \cite{8337900}. These challenges have already been identified in recent reports by the U.S. Homeland Security Committee, which discusses potential countermeasures and emphasizes the importance of developing interceptor UAVs with precise detection, tracking, and interception mechanisms as a holistic system that can effectively neutralize this threat \cite{homelandsecurity}.

In accordance, this work investigates the deployment of a team of pursuing UAVs (i.e., agents) over a confined area (sensitive facility) with the aim of searching, detecting, and tracking multiple rogue drones (i.e., targets). In particular, in this work we propose a multiagent reinforcement learning (RL) scheme for simultaneously detecting and tracking multiple rogue drones by a team of autonomous UAVs agents. We focus on a realistic scenario in which the rogue drones and the pursuer UAV agents move in a confined space, where each agent is equipped with a range-finding sensor that exhibits a limited sensing range for detecting the rogue drones. More specifically, the agents can detect the presence of the rogue drones only if they are inside their sensing range. In the proposed framework the UAVs aim at: (a) learning how to effectively be deployed and search their controlled area to accurately detect and track rogue drones, and (b) best choosing their mobility actions to maximize the time they have these targets in their field-of-view (FOV). The main contributions of this work are as follows:
\begin{itemize}
\item The search-and-track problem is formulated as a Markov Decision Process (MDP), and a novel approach based on multiagent reinforcement learning is proposed. Specifically, multiple Q-learning agents cooperate, thus allowing the team of pursuer UAV agents to coordinate and optimize their mobility control actions at each time-step in order to maximize the aggregated number of targets detected over the operating time period.
\item  Two reward functions are proposed that encourage the coordination between the learning agents to closely achieve the constraints and the objectives of the cooperative search-and-track problem. 
\item Extensive simulation experiments demonstrate the effectiveness of the proposed approach for various scenarios with changing numbers of pursuer agents.
\end{itemize}

The rest of the paper is organized as follows. Section \ref{sec:related} summarizes the related work and provides a brief introduction to RL. Section \ref{sec:system_model} presents the system model and formulates the search-and-track problem. Section \ref{sec:proposed_approach} describes the proposed RL-based multiple UAV control approach for searching-and-tracking, while Section \ref{sec:evaluation} presents detailed simulation results. Finally, Section \ref{sec:conclusions} concludes the work.

\section{Background and Related Work} \label{sec:related}

\subsection{State of the Art}
\label{ssec:sota}
Rogue drone detection and tracking technologies mainly rely on radars, RF signals, computer vision, acoustic signals, and sensor fusion \cite{Guvenc2018}. Existing anti-drone systems typically utilize fixed terrestrial sensors; for example, \cite{Multerer2017} describes a solution that uses radars, while \cite{9089515} uses RF analysis and \cite{Shi2018} fuses acoustic signals, computer vision, and RF signals. Unfortunately, ground-based solutions are subject to limitations, especially in harsh environments like cities, due to obstacles.

More recent works in~\cite{SouliICUAS2022,SouliICUAS2023} describe a prototype passive radar system, deployed on the ground and onboard a UAV. This prototype is based on signals of opportunity and software-defined radio, for detecting-and-tracking malicious drones operating over sensitive regions.

Cooperative teams of UAV agents unlock significantly more capabilities than what is possible by a single UAV. The work in \cite{9064696} proposes a dynamic radar network, composed of UAVs outfitted with radars, that is able to distributively detect-and-track rogue drones in real time. Further, the authors in \cite{9406813} propose an approach based on deep reinforcement learning for tracking multiple targets using multiple UAV agents. However, the aforementioned works only focus on target tracking and no searching for targets is performed, i.e., prior information about the targets is assumed. The authors in \cite{Yliniemi_Agogino_Tumer_2014,7759651} coordinate multiple robots to explore an area by observing multiple static points of interest (POIs) using reinforcement learning. There is no assumption of prior information about the POIs and the coordination is achieved through a novel informative reward function.

Complementing the state of the art, this work investigates the joint search-and-track problem (i.e., to locate and track rogue drones) that arises when multiple pursuer agents aim to intercept multiple rogue drones. As elaborated in the previous section, this important task precedes the track-and-intercept task that we have previously investigated in \cite{PV2023}, where multiple UAV agents track-and-intercept the operation of multiple rogue drones by transmitting jamming signals to interrupt their communication links (RF) and sensing receivers (e.g. GNSS). 

\subsection{Background on Reinforcement Learning} \label{ssec:background}
Reinforcement learning \cite{sutton1998} is a paradigm in which an agent learns to make decisions by interacting with an environment. Through trial-and-error, the agent explores various actions and receives a numerical feedback from the environment in the form of a reward that is relative to its actions. The goal of an RL problem is to find an optimal policy, i.e., a mapping from states to actions, that maximizes the cumulative rewards over time. 

Q-learning, is a widely-used algorithm in which an optimal policy is learnt by estimating the values of state-action pairs, $Q(s,a)$, and storing them in a table, referred to as the Q-table. The update rule for Q-learning is as follows:
\begin{equation} \label{eq:Q_update}
Q(s,a) \leftarrow Q(s,a)+\alpha[r+\gamma \max_{a'} Q(s',a')-Q(s,a)],
\end{equation}
\noindent where $\alpha$ is the learning rate and $r$ is the reward received for transitioning from state $s$ to state $s'$ by taking action $a$. 

Since many RL applications involve large or continuous state and action spaces, they usually require function approximation. Tile coding \cite{sutton1998} is a commonly used linear state function approximator that exhaustively partitions the state space into a set of overlapping tilings and tiles.

A cooperative multi-agent system is comprised of several decision-making agents operating in a common environment, interacting with each other to achieve a common objective \cite{wooldridge2009introduction}. This common objective is typically complex, and multi-agent RL (MARL) \cite{4445757} is a promising way to address the emerging complexity. A typical reward function in MARL problems is the {\it global reward} $G$ which represents the system's performance and motivates the agents to act in the system's interest; however, all agents receive the same reward, irrespective of whether their actions improved the system performance, leading to poor quality information (i.e., low signal-to-noise ratio). This is referred to as the credit assignment problem and can be tackled with reward shaping, which refers to a collection of methods (e.g., \cite{devlin2011theoretical, malialis2019resource, Yliniemi_Agogino_Tumer_2014}) which modify (or ``shape'') the original reward function.

In this work we have adopted the popular difference rewards method. Contrary to others, it has some theoretical properties, it does not assume any domain knowledge, and it is generic/domain-agnostic. A {\it difference reward} $D_j$ \cite{Yliniemi_Agogino_Tumer_2014} is a shaped reward signal that quantifies each agent's individual contribution to the system's performance, by removing a large amount of the noise created by the actions of other agents active in the system, and is given by:
\begin{equation} \label{eq:difference_reward}
D_j(s,a) = G(s,a)-G(s_{-j},a_{-j}),
\end{equation}
\noindent where $G(s,a)$ is the global reward and $G(s_{-j},a_{-j})$ is the counterfactual term, i.e., the global reward for a theoretical system without the contribution of the $j$-th agent.

\section{System Model} \label{sec:system_model}
\subsection{Target Dynamics}
\label{ssec:target_dynamics}
In this work, a set of rogue drones (i.e., targets), denoted by $I=\{1, 2, ..., M\}$, maneuver in the environment with discrete-time dynamics. The state vector of a target $i$ at time-step $t$ is denoted by $x^i_t=[x_\text{x}^i,\dot{x}_\text{x}^i,x_\text{y}^i,\dot{x}_\text{y}^i]^\top_t \in \mathbb{R}^4$ containing the position $(x_\text{x}^i,x_\text{y}^i)$ and velocity $(\dot{x}_\text{x}^i,\dot{x}_\text{y}^i)$ in 2D Cartesian coordinates. The targets start from a random point in the perimeter of the area and move linearly towards a POI with constant velocity magnitude.
 
\subsection{Agent Dynamics} \label{ssec:AgentDynamics}
A set of controllable UAV agents, denoted by $J=\{1, 2, ..., N\}$, cooperate to counter the operation of the rogue drones. At time-step $t$, the kinematic state of agent $j$ is denoted by  $\bar{s}^j_{t} = [\bar{s}^j_\text{x},\bar{s}^j_\text{y}]^\top_{t} \in \mathbb{R}^2$, i.e., the position in 2D Cartesian coordinates, and its dynamics are given by:
\small
\begin{equation} \label{eq:controlVectors}
\bar{s}^j_{t} = \bar{s}^j_{t-1} + a_{l_1,l_2} = \bar{s}^j_{t-1} + \begin{bmatrix}
\Delta_R[l_1] \text{cos}(l_2 \Delta_\theta)\\
\Delta_R[l_1] \text{sin}(l_2 \Delta_\theta)
\end{bmatrix},  
\begin{array}{l} 
l_1 = 1,...,|\Delta_R|\\
l_2 = 0,...,N_\theta 
\end{array} 
\end{equation}
\normalsize

\noindent where $\Delta_R$ is the vector of possible radial step sizes (with $\Delta_R[l_1]$ returning the value at index $l_1$), $\Delta_\theta=2\pi/N_\theta$, and the parameters ($|\Delta_R|$, $N_\theta$) determine the number of possible mobility control actions. The mobility control action $a_{l_1,l_2}$ represents the direction and step size of the movement along the $x$ and $y$ axis. At each time-step $t$, each UAV agent chooses one mobility control action from the discrete set $\mathbb{U}=\{a_{l_1,l_2} \mid l_1 = 1,...,|\Delta_R|, ~l_2 = 0,...,N_\theta \} $, as computed by Eq. (\ref{eq:controlVectors}), while agent $j$'s set of all admissible mobility controls is denoted by $\mathbb{U}_t^j=\{\bar{s}^{j,1}_{t},\bar{s}^{j,2}_{t},...,\bar{s}^{j,|\mathbb{U}_t^j|}_{t}\}$.  

\subsection{Agent Sensing Model} \label{ssec:agent_sensing}
The UAV agents are able to detect nearby targets only if they are inside their limited FOV. More specifically, a target at state $x^i_t$ is detected by agent at state $\bar{s}^j_t$, if the Euclidean distance between them at time-step $t$, $d^{i,j}_t$, is below the detection radius $R_0$; i.e., if $d^{i,j}_t=\norm{H x^i_t-\bar{s}^j_t}_2 \le R_0$ holds, where $H$ is a matrix that extracts the position coordinates from a target's state vector.

\subsection{Problem Formulation} \label{ssec:problem_formulation}
The objective of the search-and-track problem tackled in this work is to maximize the aggregated number of targets detected by the agents over an episode, given by:
\begin{align}
(\text{P1})& \max_{u_t^j} ~~F = \sum_{t=1}^{T} \sum_{i \in I} N^i_t \\
\mathrm{s.t.} &\>\>  u^j_t \in \mathbb{U}_t^j, \>\>\forall j \in J \>\> \forall t
\end{align}
\noindent where $F$ denotes the team's utility function, i.e., the global system performance, which is the overall team's objective to maximize. $N^i_t$ denotes whether target $i$ is being detected by an agent at time-step $t$, and is given as:
\begin{equation}  \label{eq:N^i_t}
N^i_t= 
\begin{cases} 
1,  & \!\!\text{if } \exists j \in J: d_t^{i,j} \le R_0  \\
0,  & \!\!\text{otherwise}
\end{cases} 
\end{equation}  

\noindent i.e., it is equal to $1$ if at least one agent $j \in J$ detected target $i$ at time-step $t$ and $0$ otherwise, and $d_t^{i.j}$ is the Euclidean distance between the $i$-th target and $j$-th agent (computed as given in Section \ref{ssec:agent_sensing}). Overall, the optimization problem (P1) finds the mobility control actions $u_t^j$ for each agent $j \in J$ and each time-step $t$, that maximize the sum of the number of targets detected over an episode. It should be noted that in order to optimally solve (P1), an assumption must be made about prior knowledge of the target trajectories, i.e., the target states $x^i_t$ for each time-step $t$.

\section{Proposed Approach} \label{sec:proposed_approach}
\begin{figure}
\centering
\includegraphics[width=0.7\columnwidth]{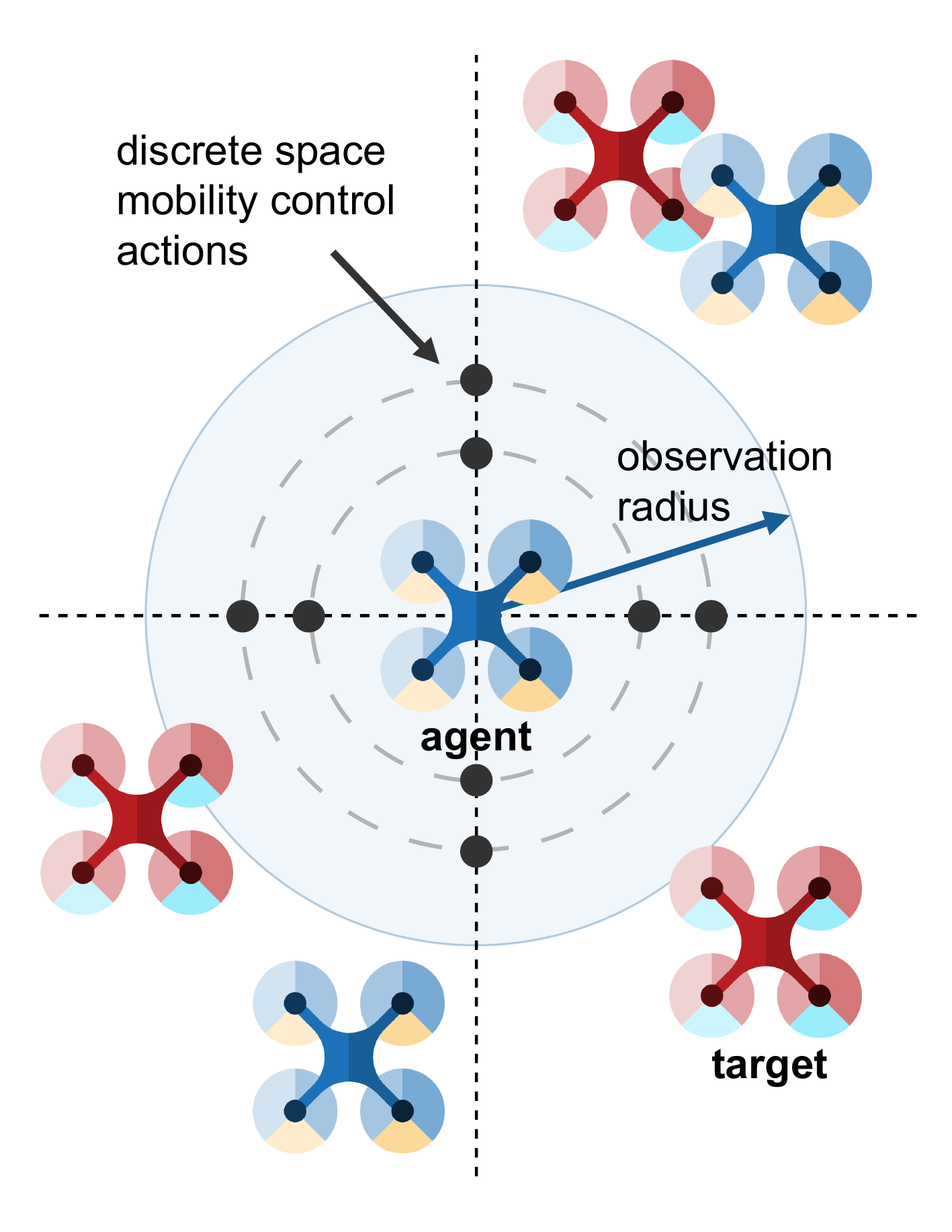}
\caption[A sample figure]{Multiple UAV agents searching and tracking multiple rogue drones\protect\footnotemark. Each agent detects targets within its detection radius in each of its four sensing areas.}
\label{fig:system_model}
\end{figure}

This section presents the RL-based approach for searching and tracking multiple targets using multiple agents. Prior knowledge about the targets, such as their locations, the number of targets, and their kinematic model is not known to the agents and the agents must coordinate to maximize the team's utility function, $F$. Figure \ref{fig:system_model} illustrates the problem tackled in this work.
\footnotetext{Drone icons made by Freepik [\url{https://www.flaticon.com/authors/freepik}] from \url{www.flaticon.com}.}

\subsection{Proposed Cooperative Q-learning Algorithm} 
\label{ssec:mdp}
In order to address the described search-and-track problem using RL, it is first formulated as an MDP \cite{sutton1998}:

\noindent \textbf{State:} The state vector of agent $j$ at time-step $t$ is determined by the positions of the agents and the position of each target $i$ (information about a target's position is available only if the target is within agent $j$'s FOV, as discussed in Section \ref{ssec:agent_sensing}), and is denoted by $s^j_t=[\bar{s}^j, m^{j1}, ...,m^{j4}, p^{j1}, ..., p^{j4}, n^{j1}, ..., n^{j4}, q^{j1}, ..., q^{j4}]_t$. The state vector of the $j$-th agent comprises five components:
\begin{itemize}
\item $\bar{s}^j$: absolute position coordinates of agent $j$
\item $m^{jl}$: number of targets in the $l$-th area of agent $j$
\item $p^{jl}$: average Euclidean distance between agent $j$ and the targets in the $l$-th area of agent $j$
\item $n^{jl}$: number of agents in the $l$-th area of agent $j$
\item $q^{jl}$: average Euclidean distance between agent $j$ and the remaining agents in the $l$-th area of agent $j$  
\end{itemize}

\noindent where $l \in \{1,2,3,4\}$. Essentially, at each time-step $t$ each agent $j$ splits the 2D space in four areas by assuming Cartesian axes centered at its current position coordinates and computes relative distances from the agents and targets. Importantly, the size of the state vector is constant and independent of the number of agents and targets and thus allows the framework to scale well with respect to the number of learning agents and targets.

\noindent \textbf{Action:} At each time-step $t$, each agent $j$ selects an action $a_t^j$ from the discrete action space $\mathcal{A}=\mathbb{U}$, which includes choosing a mobility control as defined in Section \ref{ssec:AgentDynamics}. Illegal actions are restricted, e.g., an action that moves an agent outside of the environment's area.

\noindent \textbf{Reward:} After observing its state and taking an action, each agent $j$ receives a scalar reward, $r^{j}_t$, from the environment. Various reward functions are investigated for the search-and-track problem:

The \textit{global reward, $G_t$,} is defined as the number of individual targets the team as a whole detected at time-step $t$, given by:
\begin{equation} \label{eq:G}
G_t=\sum_{i\in I} N^i_t,
\end{equation}
\noindent where $N^i_t$ is given by Eq. (\ref{eq:N^i_t}).

The \textit{difference reward, $D^j_t$,} for agent $j$ is computed by applying Eq. (\ref{eq:difference_reward}):
\begin{equation}\label{eq:D^j}
D_{t}^j= G_{t} - G_{t}^{(-j)},
\end{equation}
\noindent where the counterfactual term, $G_{t}^{(-j)}$, is calculated by assuming that agent $j$ does not detect any target for time-step $t$, as given by:
\begin{equation}\label{eq:G_-j}
G_{t}^{(-j)} =  \sum_{i\in I} \tilde{N}^i_t, ~~~\tilde{N}^i_t= 
\begin{cases} 
1,  & \!\!\text{if } \exists j' \in J, j' \ne j: d_t^{i,j'} \le R_0  \\
0,  & \!\!\text{otherwise.}
\end{cases}
\end{equation} 

The RL-based training algorithm is summarized in Alg. \ref{alg:Algorithm1}. Each UAV agent is an independent learner, having its own Q-table which is initialized with zero values (line 1). For every episode, which lasts for $T$ consecutive time-steps, the environment is initialized randomly, encouraging the agents to learn various strategies via trial-and-error (line 3). Each agent selects an action according to an $\epsilon$-greedy policy, i.e., selects an action that corresponds to the maximum Q-value in the current state with probability $1-\epsilon$ or to a random action with probability $\epsilon$ (line 6). The exploration parameter $\epsilon$ and learning rate $\alpha$ decay over time in order to encourage learning and exploration at initial episodes and exploitation afterwards (lines 12-13). Each agent executes its action, receives the reward, and observes the new state before updating the Q-value of the current state-action pair (lines 7-9).
\begin{algorithm}
\caption{Q-learning training for multi-UAV searching and tracking of rogue drones.}
\small
\label{alg:Algorithm1}
\begin{algorithmic}[1]
\STATE Initialize Q-tables for each agent with zero values
\FOR {$episode$ $n=1,2,...,N_e$}
\STATE Initialize environment
\FOR {$timestep$ $t=1,2,...,T$}
\FOR {$agent$ $j=1,2,...,N$}
\STATE Choose a greedy action $a^j_t=\text{argmax}_{a_t \in \mathcal{A}}Q^j(s^j_t,a_t)$ with probability $1-\epsilon$ or a random action with probability $\epsilon$
\STATE Take action $a^j_t$
\STATE Observe reward $r^j_t$ and next state $s^j_{t+1}$
\STATE Update $Q^j(s^j_t,a^j_t)$ using Eq. (\ref{eq:Q_update})
\ENDFOR
\ENDFOR
\STATE Decay $\alpha$ using $alpha\_decay\_rate$
\STATE Decay $\epsilon$ using $epsilon\_decay\_rate$
\ENDFOR
\end{algorithmic}
\end{algorithm}

\normalsize
\section{Performance Evaluation} \label{sec:evaluation}
\subsection{Simulation Setup}
\label{sec:simulation_setup}
\begin{figure} 
\includegraphics[width=\linewidth]{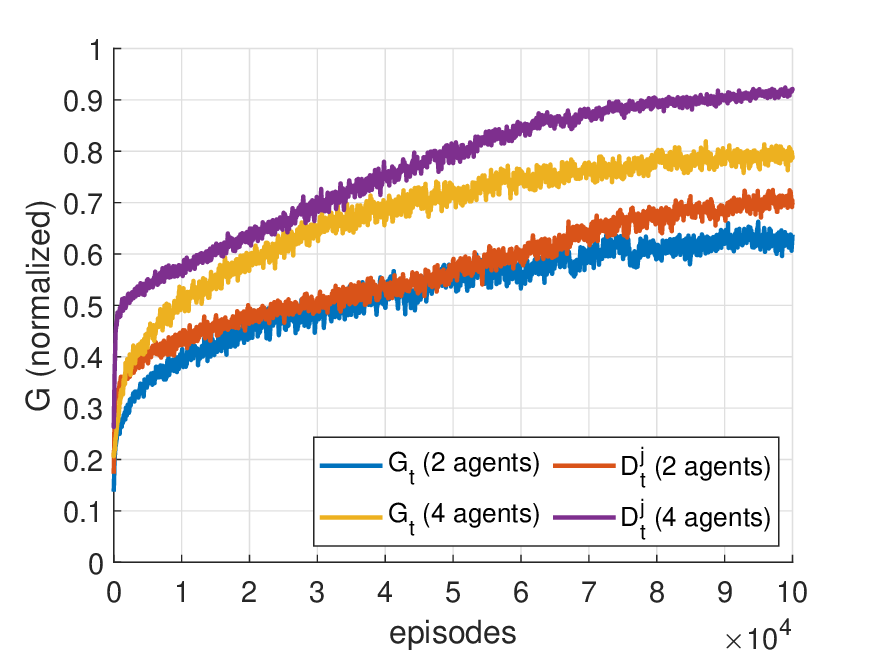}  \caption{Training results for different number of agents and reward functions.}
\label{fig:training_curves}
\end{figure}
\begin{figure*}
\begin{subfigure}{\linewidth}  
\includegraphics[width=0.7\linewidth]{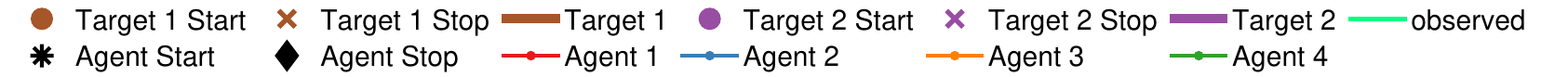} 
\end{subfigure}
\begin{subfigure}{\linewidth}
\begin{subfigure}{0.329\linewidth}
\centering  
\includegraphics[width=\linewidth]{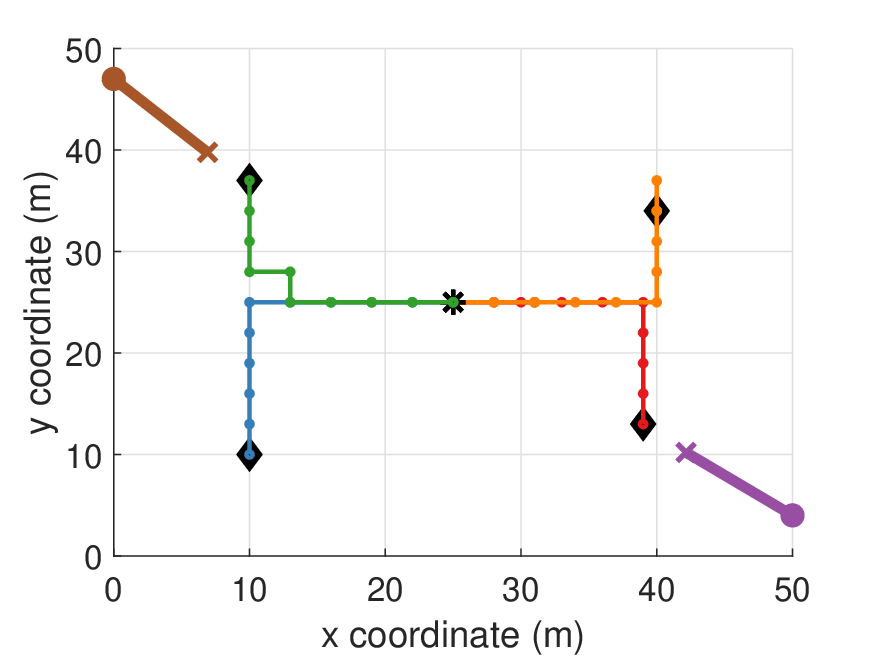}  
\caption{$t=1,...,10$ s} \label{fig:scenario_1}  
\end{subfigure}
\begin{subfigure}{0.329\linewidth}
\centering  
\includegraphics[width=\linewidth]{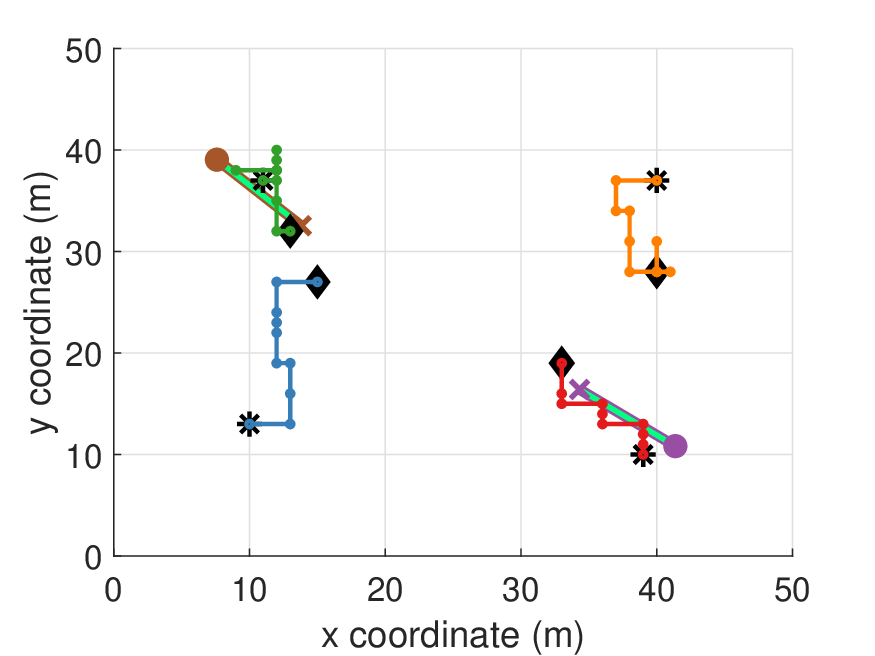}   
\caption{$t=11,...,20$ s} \label{fig:scenario_2} 
\end{subfigure}
\begin{subfigure}{0.329\linewidth}
\centering  
\includegraphics[width=\linewidth]{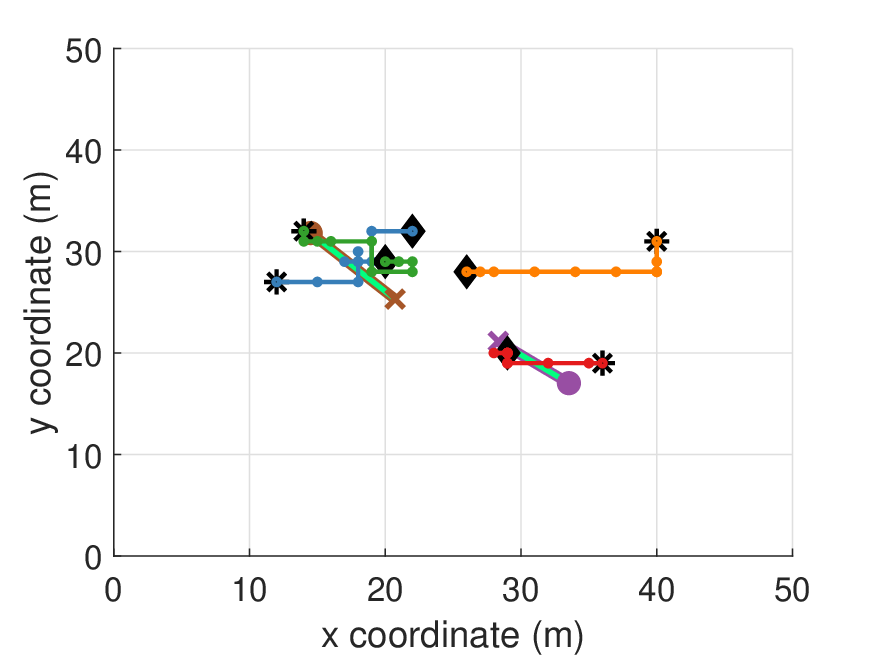}   
\caption{$t=21,...,30$ s} \label{fig:scenario_3} 
\end{subfigure}
\end{subfigure}
\caption{Maneuvers of $4$ search-and-track agents in a simulated scenario with $2$ targets, during $3$ time periods (agent start and stop positions are marked with $\star$ and $\blacklozenge$, respectively, and target start and stop positions are marked with $\bullet$ and $\times$, respectively).}
\label{fig:sim_scenario}
\end{figure*}
\begin{figure*}
\begin{subfigure}{\linewidth}
\begin{subfigure}{0.329\linewidth}
\centering  
\includegraphics[width=\linewidth]{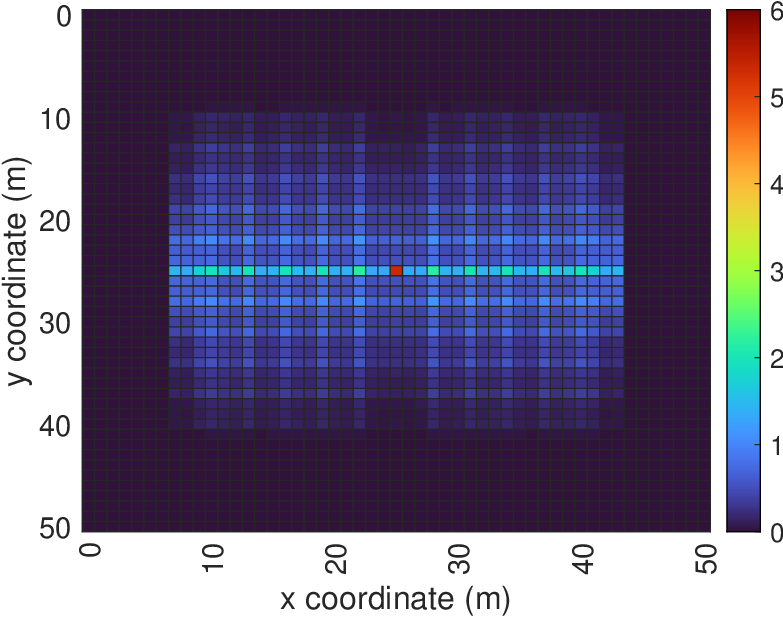}   
\caption{(P1)} \label{fig:heatmap_opt_4a} 
\end{subfigure}
\begin{subfigure}{0.329\linewidth}
\centering  
\includegraphics[width=\linewidth]{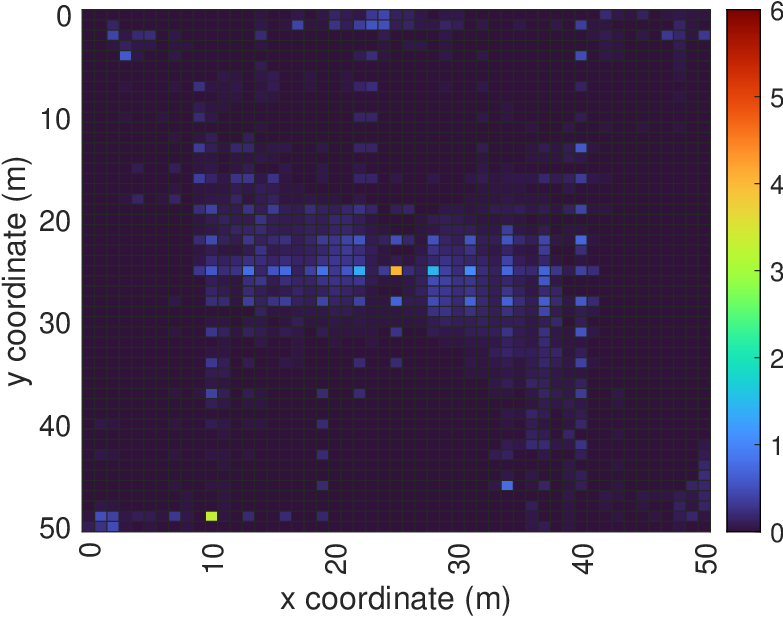}  
\caption{RL ($r^j_t=G_t$)} \label{fig:heatmap_G_4a}  
\end{subfigure}
\begin{subfigure}{0.329\linewidth}
\centering  
\includegraphics[width=\linewidth]{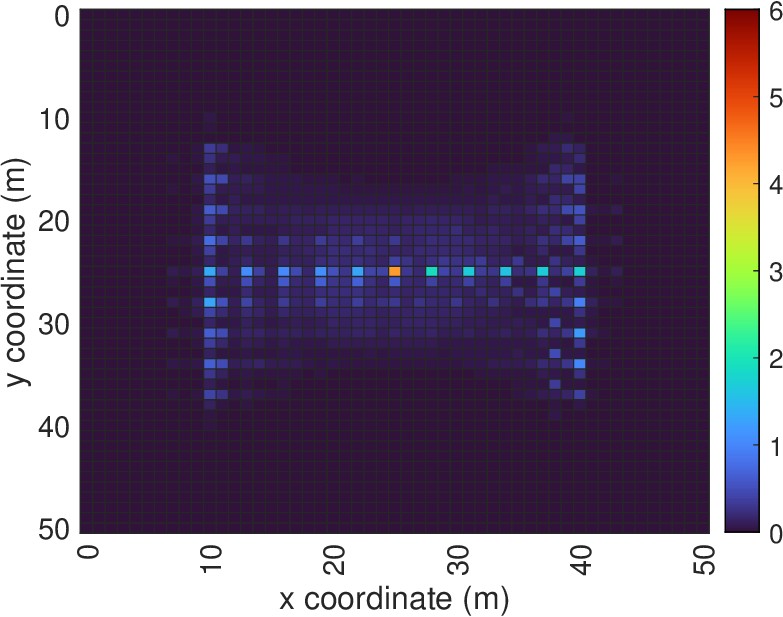}   
\caption{RL ($r^j_t=D^j_t$)} \label{fig:heatmap_D_4a} 
\end{subfigure}
\end{subfigure}
\caption{Optimal (P1), and learned ($G$, $D$) agent trajectories (covering the area under consideration) averaged over all training episodes (visits per square meter) depicted for the case of $4$ pursuing agents.}
\label{fig:heatmaps}
\end{figure*}
\noindent \textbf{Simulations:} The UAV agents and the rogue drones operate in an area of size $50$ m by $50$ m. The targets are randomly initialized in the perimeter of the area where: target $1$ is initialized with $x^1_y$ ranging from $0$ to $50$ m and $x^1_x=0$ m, and target $2$ is initialized with $x^2_y$ ranging form $0$ to $50$ m and $x^2_x=50$ m. The agents are initially positioned in the middle of the area, i.e., at $[25,25]^\top$ m. The agent's dynamical model has radial displacement $\Delta_R=[0,1,3]$ m, and $N_\theta =4$ which gives a total of $9$ control actions, including the initial position of the agent. The POI coordinates are random inside a $10$ m $\times 10$ m box in the middle of the area. An agent can detect a target only if the target is within distance $R_0=5$ m from the agent. 

\noindent \textbf{RL:} The agents learn for $N_e=10^5$ episodes, where each episode is comprised of $T=30$ time-steps. The Q-learning parameters are set as follows: the learning rate is set to $\alpha=0.2$ and $alpha\_decay\_rate=0.99997$, and the exploration parameter is set to $\epsilon=0.3$ and $epsilon\_decay\_rate=0.99997$, while the discount factor is set to $\gamma = 0.9$. For representing the state space, each agent has its own tile coding function approximator comprising $2^6$ grid tilings. The tile width in $m^{jl}$ and $n^{jl}$ dimensions is set to $1$, while in $\bar{s}^j$, $p^{jl}$ and $q^{jl}$ dimensions is set to $25$ m. The values for the Q-learning and tile coding parameters are selected after performing parameter tuning.  

\noindent \textbf{Evaluation:} The experiments are repeated over $20$ independent runs. All training curve plots show average values across $20$ independent runs with a median filter applied.

\subsection{Simulation Results} 
We begin the evaluation of the proposed approach with a comparison between the learning performance of the proposed reward functions (discussed in Section \ref{ssec:mdp}) when utilizing $2$ and $4$ agents. Figure \ref{fig:training_curves} shows the global system performance normalized in the range $[0,1]$. As observed in Fig. \ref{fig:training_curves}, the learning patterns improve as the training episodes progress for both reward functions. However, the difference reward, $D^j_t$, outperforms the global reward, $G_t$, in terms of the learning speed, scalability, and overall performance, with its advantage being more evident in the $4$-agent case, since the higher number of agents increases the noise created by the other agents active in the system, making it more difficult to learn from $G_t$. At the last training episode, $D_t^j$ converges to  approximately $0.9$ which is very close to the optimal value, i.e., $1$. The optimal value, however, requires prior knowledge about the targets (as discussed in Section \ref{ssec:problem_formulation}). Even so, due to its higher performance, the experiments that follow, utilize the difference reward, $D^j_t$, as the reward function $r^j_t$.

Figure \ref{fig:sim_scenario} shows a simulated scenario of the policy learned with $4$ agents and $2$ targets during $3$ consecutive time periods, i.e. $t=1,...,10$ s, $t=11,...,20$ s, $t=21,...,30$ s. The agents' positions are initialized in the middle of the area, i.e., at $[25,25]^\top$ m. The initial target state vectors are $[0,0.69,47,-0.72]^\top$ and $[50,-0.78,4,0.62]^\top$ for targets $1$ and $2$ (in units m, m/s, m, and m/s, respectively). The POIs are located at $[21,25]^\top$ m and $[28,21]^\top$ m for targets $1$ and $2$, respectively. The trajectories of the agents and the targets are shown in Figs. \ref{fig:scenario_1} - \ref{fig:scenario_3}, for the three time periods, where the bright green line indicates whether a target is detected. During the first time period, the $4$ agents spread out towards $4$ different directions to maximize the coverage of the area and reduce the overlap in their sensing range and therefore maximize the probability of detecting the targets. At the beginning of the second time period, at time-step $t=11$ s, agents $4$ and $1$ detect targets $1$ and $2$, respectively, and continue to detect them by tracking them for the rest of the period. The remaining agents, i.e., agents $2$ and $3$, maneuver towards the center of the area to search for targets, since during this time period it is more likely to find a target there. During the last time period, both targets continue being tracked by the agents. 

To further evaluate the performance of the proposed approach, search patterns are generated by running $1000$ episodes under the agents’ learned policy and  over $20$ independent runs. Figure \ref{fig:heatmaps} depicts the agent's search-and-track patterns by illustrating a heat-map of the agent trajectories across the whole search area. To create these heat-maps, the 2D area was split into tiles of square meter each and we recorded the frequency that each tile was traversed by each agent across all episodes and runs for both reward functions $G_t$, $D^j_t$, as well as by optimally solving problem (P1). As observed from Fig. \ref{fig:heatmaps}, the agents search the area more effectively when using the difference reward, $D^j_t$, as compared to the global reward, $G_t$, resembling closer the trajectories obtained utilizing the solutions obtained by (P1). Finally, Table \ref{table:1} reports on the achieved performance for both reward functions and for (P1), along with the normalized values in the parenthesis, for utilizing $2$ and $4$ agents. The results verify the advantage of the difference reward over the global reward, which for the case of $4$ agents reaches the normalized value of $94.75\%$.
\begin{table}
\centering
\setlength{\tabcolsep}{3pt}
\caption{Comparison of the two reward functions used to optimally solve (P1) (that assumes prior target knowledge).}
\begin{tabular}{cc|ccc}
\hline 
\multirow{3}{*}{\begin{tabular}[c]{@{}c@{}}Performance metric\\($F$)\end{tabular}} &      & \multicolumn{3}{c}{Method} \\ \cline{3-5} 
&      & \multicolumn{2}{c|}{RL-based}          & \multirow{2}{*}{(P1)} \\ \cline{3-4}
&  & \multicolumn{1}{c|}{$r^j_t=G_t$}           & \multicolumn{1}{c|}{$r^j_t=D^j_t$}           & \\ \hline
\multirow{2}{*}{$2$ agents} & mean & \multicolumn{1}{c|}{30.28 (64.88\%)} & \multicolumn{1}{c|}{33.96 (72.77\%)} & 46.67 \\
& std  & \multicolumn{1}{c|}{2.69} & \multicolumn{1}{c|}{2.27} & 0.45  \\ \hline
\multirow{2}{*}{$4$ agents}  & mean & \multicolumn{1}{c|}{37.55 (80.49\%)} & \multicolumn{1}{c|}{44.20 (94.75\%)} & 46.65 \\
& std  & \multicolumn{1}{c|}{1.98} & \multicolumn{1}{c|}{0.86} & 0.44 \\ \hline
\end{tabular}
\label{table:1}
\end{table}

\section{Conclusions} \label{sec:conclusions}
This work has examined the utilization of multiple UAV agents to cooperatively search-and-track multiple rogue drones. We have proposed an RL-based scheme, that utilizes a simple reward function which encourages the coordination between the learning agents. The learned strategies enable the agents to select mobility control actions for each agent to maximize the aggregated number of targets detected over each episode, considering the agents' limited sensing range for detecting the targets. Performance evaluation results have provided a comparison between two reward functions (global and difference rewards) and demonstrated the superiority of the difference reward function in terms of scalability, learning speed, and final joint performance. 


As future work we plan to integrate these results with our previous track-and-jam work to provide a holistic search-track-jam system for rogue-drone interception. Moreover we aim at field-testing these algorithms in a small scale deployment with commercial-off-the-shelf drones using the onboard LiDAR ranging sensors and software-defined-radios as payloads to enable jamming. The goal is to execute the integrated search-track-jam scheme to test its real-time performance.

\bibliographystyle{IEEEtran}
\bibliography{IEEEabrv,main}

\begin{thebibliography}{10}
\providecommand{\url}[1]{#1}
\csname url@samestyle\endcsname
\providecommand{\newblock}{\relax}
\providecommand{\bibinfo}[2]{#2}
\providecommand{\BIBentrySTDinterwordspacing}{\spaceskip=0pt\relax}
\providecommand{\BIBentryALTinterwordstretchfactor}{4}
\providecommand{\BIBentryALTinterwordspacing}{\spaceskip=\fontdimen2\font plus
\BIBentryALTinterwordstretchfactor\fontdimen3\font minus \fontdimen4\font\relax}
\providecommand{\BIBforeignlanguage}[2]{{%
\expandafter\ifx\csname l@#1\endcsname\relax
\typeout{** WARNING: IEEEtran.bst: No hyphenation pattern has been}%
\typeout{** loaded for the language `#1'. Using the pattern for}%
\typeout{** the default language instead.}%
\else
\language=\csname l@#1\endcsname
\fi
#2}}
\providecommand{\BIBdecl}{\relax}
\BIBdecl

\bibitem{9194729}
H.~Kang, J.~Joung, J.~Kim, J.~Kang, and Y.~S. Cho, ``Protect your sky: A survey of counter unmanned aerial vehicle systems,'' \emph{IEEE Access}, vol.~8, pp. 168\,671--168\,710, 2020.

\bibitem{8678424}
D.~Schneider, ``Regulators seek ways to down rogue drones: Growing antidrone industry offers radar, remote {ID}, and other tools,'' \emph{IEEE Spectrum}, vol.~56, no.~4, pp. 10--11, 2019.

\bibitem{hackingdrones}
K.~Wesson and T.~Humphreys, ``Hacking drones,'' \emph{Scientific American}, vol. 309, no.~5, pp. 54--59, 2013.

\bibitem{9378538}
S.~Park, H.~T. Kim, S.~Lee, H.~Joo, and H.~Kim, ``Survey on anti-drone systems: Components, designs, and challenges,'' \emph{IEEE Access}, vol.~9, pp. 42\,635--42\,659, 2021.

\bibitem{FAA}
\BIBentryALTinterwordspacing
M.~O'donnell, ``{Letter from FAA office of airports on guidance on use of counter UAS systems at airports},'' \emph{{US FAA}}, 2018. [Online]. Available: \url{https://www.faa.gov/airports/airport safety/media/Attachment-1-Counter-UAS-Airport-Sponsor-Letter-July-2018.pdf}
\BIBentrySTDinterwordspacing

\bibitem{9765451}
M.~A. Khan, H.~Menouar, A.~Eldeeb, A.~Abu-Dayya, and F.~D. Salim, ``On the detection of unauthorized drones—techniques and future perspectives: {A} review,'' \emph{{IEEE Sensors Journal}}, vol.~22, no.~12, pp. 11\,439--11\,455, 2022.

\bibitem{8337900}
I.~Guvenc, F.~Koohifar, S.~Singh, M.~L. Sichitiu, and D.~Matolak, ``Detection, tracking, and interdiction for amateur drones,'' \emph{IEEE Communications Magazine}, vol.~56, no.~4, pp. 75--81, 2018.

\bibitem{homelandsecurity}
T.~Humphreys, ``Statement on the security threat posed by unmanned aerial systems and possible countermeasures,'' \emph{Oversight and Management Efficiency Subcommittee, Homeland Security Committee}, 2015.

\bibitem{Guvenc2018}
I.~{Guvenc}, F.~{Koohifar}, S.~{Singh}, M.~L. {Sichitiu}, and D.~{Matolak}, ``Detection, tracking, and interdiction for amateur drones,'' \emph{IEEE Commun. Mag.}, vol.~56, no.~4, pp. 75--81, 2018.

\bibitem{Multerer2017}
T.~{Multerer}, A.~{Ganis}, U.~{Prechtel}, E.~{Miralles}, A.~{Meusling}, J.~{Mietzner}, M.~{Vossiek}, M.~{Loghi}, and V.~{Ziegler}, ``Low-cost jamming system against small drones using a {3D MIMO} radar based tracking,'' in \emph{Proc. European Radar Conf. (EURAD)}, 2017, pp. 299--302.

\bibitem{9089515}
A.~H. Abunada, A.~Y. Osman, A.~Khandakar, M.~E.~H. Chowdhury, T.~Khattab, and F.~Touati, ``Design and implementation of a {RF} based anti-drone system,'' in \emph{Proc. IEEE Inter. Conf. on Informatics, IoT, and Enabling Technologies (ICIoT)}, 2020, pp. 35--42.

\bibitem{Shi2018}
X.~{Shi}, C.~{Yang}, W.~{Xie}, C.~{Liang}, Z.~{Shi}, and J.~{Chen}, ``Anti-drone system with multiple surveillance technologies: Architecture, implementation, and challenges,'' \emph{IEEE Commun. Mag.}, vol.~56, no.~4, pp. 68--74, April 2018.

\bibitem{SouliICUAS2022}
{N. Souli, I. Theodorou, P. Kolios, and G. Ellinas}, ``Detection and tracking of rogue {UASs} using a novel real-time passive radar system,'' in \emph{Proc. Inter. Conf. on Unmanned Aircraft Systems (ICUAS)}, 2022, pp. 576--582.

\bibitem{SouliICUAS2023}
{N. Souli, P. Kolios, and G. Ellinas}, ``Onboard passive radar for detection and tracking of rogue {UAVs},'' in \emph{Proc. Inter. Conf. on Unmanned Aircraft Systems (ICUAS)}, 2023, pp. 820--826.

\bibitem{9064696}
A.~Guerra, D.~Dardari, and P.~M. Djuric, ``Dynamic radar networks of {UAVs: A} tutorial overview and tracking performance comparison with terrestrial radar networks,'' \emph{IEEE Vehicular Technology Magazine}, vol.~15, no.~2, pp. 113--120, 2020.

\bibitem{9406813}
J.~Moon, S.~Papaioannou, C.~Laoudias, P.~Kolios, and S.~Kim, ``Deep reinforcement learning multi-{UAV} trajectory control for target tracking,'' \emph{IEEE Internet of Things Journal}, vol.~8, no.~20, pp. 15\,441--15\,455, 2021.

\bibitem{Yliniemi_Agogino_Tumer_2014}
L.~Yliniemi, A.~K. Agogino, and K.~Tumer, ``Multirobot coordination for space exploration,'' \emph{AI Mag.}, vol.~35, no.~4, pp. 61--74, 2014.

\bibitem{7759651}
A.~Rahmattalabi, J.~J. Chung, M.~Colby, and K.~Tumer, ``D++: Structural credit assignment in tightly coupled multiagent domains,'' in \emph{Proc. IEEE/RSJ International Conference on Intelligent Robots and Systems (IROS)}, 2016, pp. 4424--4429.

\bibitem{PV2023}
P.~Valianti, K.~Malialis, P.~Kolios, and G.~Ellinas, ``Multi-agent reinforcement learning for multiple rogue drone interception,'' in \emph{Proc. IEEE Int. Conf. on Unmanned Aircraft Systems (ICUAS)}, 2023, pp. 1037--1044.

\bibitem{sutton1998}
R.~S. Sutton and A.~G. Barto, \emph{Introduction to Reinforcement Learning}.\hskip 1em plus 0.5em minus 0.4em\relax MIT Press Cambridge, 1998.

\bibitem{wooldridge2009introduction}
M.~Wooldridge, \emph{An Introduction to Multiagent Systems}.\hskip 1em plus 0.5em minus 0.4em\relax Wiley, 2009.

\bibitem{4445757}
L.~Busoniu, R.~Babuska, and B.~De~Schutter, ``A comprehensive survey of multiagent reinforcement learning,'' \emph{IEEE Trans. Syst. Man Cybern., Part C (Applications and Reviews)}, vol.~38, no.~2, pp. 156--172, 2008.

\bibitem{devlin2011theoretical}
S.~Devlin and D.~Kudenko, ``Theoretical considerations of potential-based reward shaping for multi-agent systems,'' in \emph{Proc. ACM Int. Conf. on Autonomous Agents and Multi-Agent Systems}, 2011, pp. 225--232.

\bibitem{malialis2019resource}
K.~Malialis, S.~Devlin, and D.~Kudenko, ``Resource abstraction for reinforcement learning in multiagent congestion problems,'' in \emph{Proc. Int. Conf. on Autonomous Agents and Multiagent Systems (AAMAS)}, 2016.

\end{thebibliography}

\end{document}